\documentclass[a4paper,twoside]{article}
\newcommand{\affil}[1]{$^{\rm #1}$}
\usepackage[authoryear]{natbib}
\bibpunct{(}{)}{;}{a}{}{,}
\usepackage{graphicx}
\date{} %Please leave the date blank
\title{\large\bf\flushleft Luminosity bias: from haloes to galaxies}
\author{\parbox{\textwidth}{\flushleft
\vspace{-0.5cm}
{\it C. M. Baugh\affil{A,B}}\\
\vspace{0.4cm}
{\small \affil{A}\,Institute for Computational Cosmology, Department of Physics, 
Durham University, South Road, Durham, DH1 3LE, UK.}\\
{\small \affil{B}\,Email: c.m.baugh@durham.ac.uk}}}
\begin{document}
\twocolumn[
\begin{changemargin}{.8cm}{.5cm}
\begin{minipage}{.9\textwidth}
\vspace{-1cm}
\maketitle
%
%
%%%%%%%%%%%%%     ABSTRACT    %%%%%%%%%%%%%
%Abstract of no more than 200 words here.
\small{\bf Abstract:}
Large surveys of the local Universe have shown 
that galaxies with different intrinsic properties, 
such as colour, luminosity and morphological 
type display a range of clustering amplitudes.
Galaxies are therefore not faithful tracers of 
the underlying matter distribution. This modulation 
of galaxy clustering, called bias, contains information 
about the physics behind galaxy formation. It is 
also a systematic to be overcome before the 
large-scale structure of the Universe can be used as 
a cosmological probe. Two types of approaches have been 
developed to model the clustering of galaxies. The first 
class is empirical and filters or weights 
the distribution of dark matter to reproduce the 
measured clustering. In the second approach an 
attempt is made to model the physics which governs  
fate of baryons in order to predict the number of 
galaxies in dark matter haloes. I will review the 
development of both approaches and summarize what 
we have learnt about galaxy bias.   

%%%%%%%%%%%%%     KEYWORDS    %%%%%%%%%%%%%
\medskip{\bf Keywords:} Write keywords here
% Please write all keywords in lower case. PASA uses the
% standard list of subject headings adopted by The Astrophysical Journal
% and available from http://www.journals.uchicago.edu/ApJ/keywords_text.html.
% Keywords are separated by em-dashes, i.e. ---

%%%%%%%%DO NOT EDIT%%%%%%%%%%%%
\medskip
\medskip
\end{minipage}
\end{changemargin}
]
\small
%%%%%%%%EDIT FROM HERE%%%%%%%%%%%%

\begin{figure*}
\begin{center}
\includegraphics[bb=200 70 560 560, scale=0.7, angle=90]{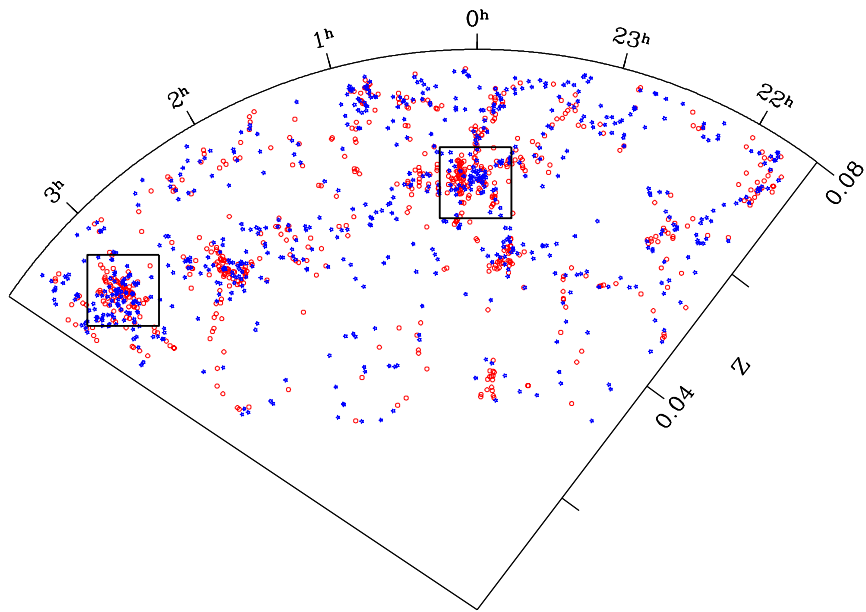}
\caption{The distribution of galaxies with early (red points) and 
late spectral (blue points) types in a volume limited sample 
(just faintwards of $L_*$), drawn from the two-degree field 
galaxy redshift survey. 
The early and late type galaxies trace out 
different features of the cosmic web.
Adapted from \cite{Norberg:2002}. 
}
\label{fig:2df}
\end{center}
\end{figure*}

\section{Introduction}

It has long been known that the distribution of galaxies on the 
sky is clumpy rather than random. Huge surveys of galaxies in the 
local Universe have further revealed that different types of 
galaxies are clustered in different ways. If galaxies are grouped 
into samples according to intrinsic properties such as their 
luminosity, colour or morphology, then the measured clustering 
varies depending on the characteristics of the galaxies under 
consideration \citep{Norberg:2001,Norberg:2002,Zehavi:2002,Zehavi:2011}. 
Fig.~\ref{fig:2df} shows this for galaxies from the two-degree 
field galaxy redshift survey which have been divided into two classes 
according to their spectral type. Galaxies with ``early'' 
or ``passive'' spectral types trace out a different pattern of 
large scale structure than the galaxies with ``late'' or ``active'' 
types. The early types delineate tighter filaments and the cores 
of clusters, whereas the late types sample the outer parts of these 
structures and appear more diffuse. 

Such differences are driven by the variation in the processes which 
shape the formation and evolution of galaxies with environment and halo 
mass. The fact that the clustering patterns of different kinds of galaxies 
look different implies that measurements of galaxy clustering have 
the potential to tell us something useful about the nature and strength 
of these processes. To realize this, we need theoreticals models which 
can describe the large-scale structure in the galaxy distribution and 
connect this to the underlying physics. 

The large-scale structure of the galaxy distribution is also used to 
constrain the values the basic cosmological parameters, including the 
equation of state of the dark energy. The distortion of the clustering 
signal due to the gravitationally induced peculiar motions of 
galaxies provides a measurement of the rate at which structure 
is growing, which in turn depends on the cosmic expansion history 
\citep{Guzzo:2008,Wang:2008}. 
The apparent location of baryonic acoustic oscillation (BAO) 
features in the power spectrum or correlation function provides 
a geometrical test, measuring the redshift-distance relation 
\citep{percival:2007,Cabre:2009,Sanchez:2009,Sanchez:2012}. The power of 
large-scale structure probes depends on how well we can model 
galaxy bias. For example, in BAO studies, the measured power 
spectrum is often divided by a featureless reference spectrum to 
remove the overall shape of the spectrum from the analysis. 
However, this shape contains further cosmological information 
if we can predict the form of the galaxy bias, so that we 
can infer the shape of the matter power spectrum. Galaxy 
bias is therefore a ``nuisance'' parameter or systematic in 
large-scale structure probes. If we can model bias, we can 
enhance the scientific performance of wide-field galaxy surveys 
by marginalizing over this parameter.  
  
In this article I will first review empirical approaches to 
modelling galaxy clustering, explaining how these developed 
as the quality of N-body simulations of hierarchical clustering 
of the dark matter improved. In the second half I will discuss 
physical approaches to predicting galaxy bias and give an overview of 
what such models have told us.

\section{Empirical models of galaxy clustering}

\begin{figure}[h]
\begin{center}
\includegraphics*[trim = 25 75 330 420, scale=0.8, angle=0]{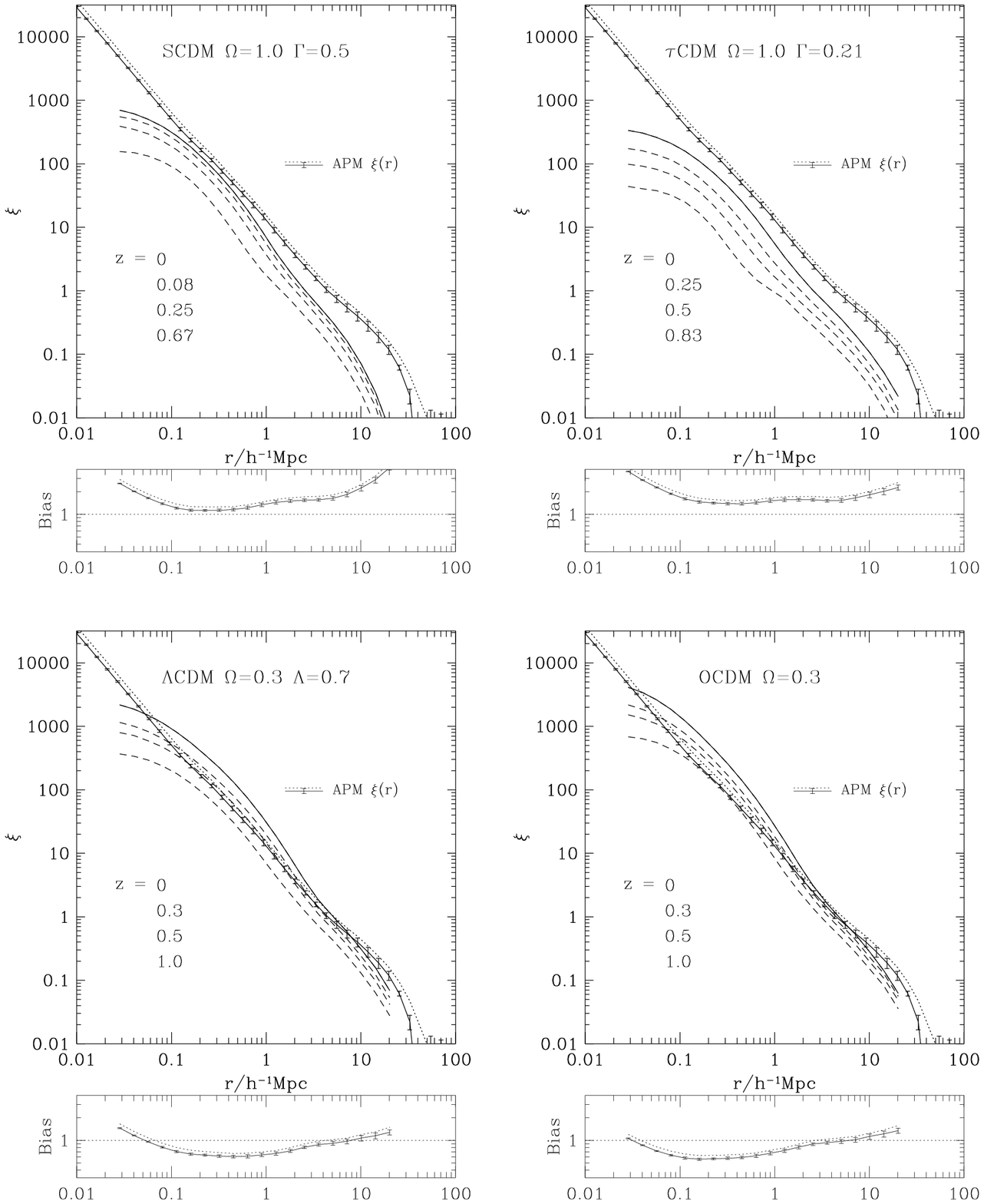}
\caption{The clustering in the matter distribution, as quantified 
through the two-point correlation function. The lines show 
measurements from N-body simulations of a $\Lambda$CDM cosmology 
at different epochs, with the upper-most curve corresponding 
to the present day. The points show a measurement of the galaxy 
correlation function, which unlike the dark matter, is well 
described by a power-law in pair separation. The effective galaxy 
bias, the square root of the ratio of the galaxy and matter 
correlation functions, is shown in the lower panel and is scale dependent.
Based on a figure from \cite{Jenkins:1998}.}
\label{fig:xiapm}
\end{center}
\end{figure}

The central pillar of the paradigm for the 
large-scale structure of the universe is 
gravitational instability. Small 
perturbations in the matter density seeded 
during inflation are amplified by gravitational 
instability. The early stages of this process 
can be followed using perturbation theory 
\citep{Bernardeau:2002}. Unless specialized 
assumptions are made, the latter, nonlinear 
stages of structure formation can only be 
modelled through numerical simulation \citep{Davis:1985}. 

N-body simulations of the hierarchical growth of 
perturbations in the density of the Universe have 
played a central role in shaping the current 
cosmological model \citep{Springel:2006}. 
According to these calculations, the correlation function of 
the dark matter at the present day cannot be 
described by a simple power law. The correlation 
function of the mass today in a cold dark matter 
universe with a cosmological constant is plotted in 
Fig.~\ref{fig:xiapm}. The correlation function of 
galaxies in a flux limited survey, roughly the 
clustering of $L_*$ galaxies, is also shown for 
contrast \citep{Baugh:1996}. In this case, the 
correlation function is impressively close to a 
power-law over more than three decades in pair 
separation. The effective galaxy bias, 
defined as the square root of the ratio of the 
galaxy and dark matter correlation functions, is 
therefore scale dependent.

Early N-body simulations lacked the resolution to 
reveal any irregularities in the structure of dark 
matter haloes. Large volume simulations suitable for 
following fluctuations on scales of tens of megaparsec 
were only able to resolve halos of group and cluster mass. 
Motivated by analytic calculations which explained the 
large correlation lengths of galaxy groups through the 
clustering of high peaks in a Gaussian density field 
\citep{Kaiser:1984}, the first attempts to model the 
spatial distribution of galaxies used the smoothed density 
field of the dark matter \citep{Davis:1985,White:1987}. 
\cite{Cole:1997} assumed that the probability of finding 
a galaxy was some empirical function of the smoothed 
density field, with parameters tuned to reproduce the 
galaxy correlation function. This approach has continued 
to be developed, with the introduction of the idea 
of stochastic bias \citep{Dekel:1999} in which the 
overdensity in the galaxy distribution can be written 
as a nonlinear function of the overdensity in the 
matter distribution with a scatter. This framework has 
been further developed and applied to surveys by a 
number of authors \citep{Sigad:2000, Szapudi:2004,Marinoni:2005,
Kovac:2011}.

\begin{figure}[h]
\begin{center}
\includegraphics[scale=0.45, angle=0]{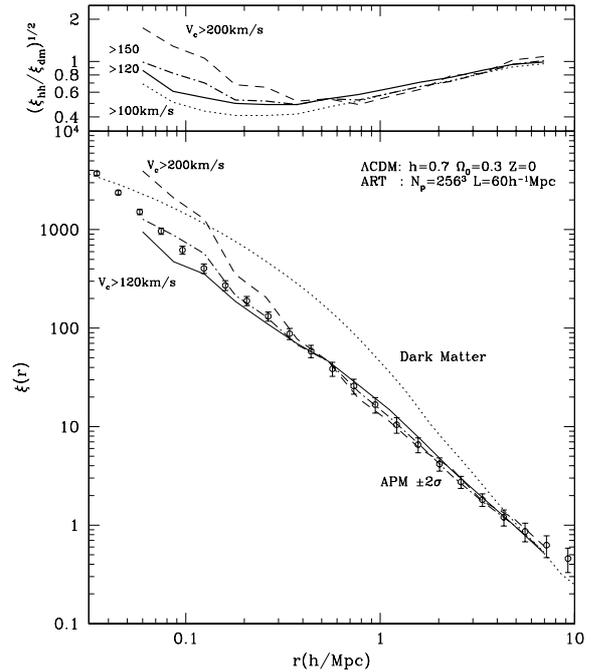}
\caption{An attempt to reproduce the observed clustering 
of galaxies by associating galaxies with subhaloes with 
effective circular velocities above some threshold value 
(dashed, dot-dashed and solid). The clustering of 
subhaloes is different from that of the overall dark 
matter (shown by the dotted line), and by tuning the circular 
velocity which defines the sample, a good match can be 
obtained with the observed galaxy clustering (shown by 
the points). Reproduced from \cite{Colin:1999}. 
}
\label{fig:xisham}
\end{center}
\end{figure}

As codes became more efficient at calculating the gravitational 
forces between large numbers of particles and the processing 
speed of computers increased, it became possible to resolve 
haloes approaching galactic masses. The clustering of haloes  
depends, in the first approximation, on halo mass, with cluster-mass 
haloes being much more strongly clustered than halos which might 
host the Milky Way \citep{Cole:1989,Mo:1996}. This led to models in which 
the form of the measured galaxy clustering could be obtained 
by applying a suitable weighting to halos, which varies with 
halo mass \citep{Jing:1998}. This is the forerunner of today's 
halo occupation distribution models in which the weighting 
is expressed in terms of the mean number of galaxies per halo, 
as described later.  

With further improvements to the simulations, it became possible 
to resolve structure inside dark matter haloes \citep{Klypin:1999,Moore:1999}. 
Haloes form through mergers and the accretion of mass. 
With sufficient resolution, the central regions of the accreted 
haloes can be preserved for many orbits, whilst 
the outer parts are stripped off. This prompted a 
new generation of modelling in which resolved subhaloes 
were associated with galaxies. In an early example of what 
today would be called ``sub halo abundance matching'', \cite{Colin:1999} 
were able to match the observed power law clustering 
of galaxies by selecting all subhalos above some threshold 
circular velocity (see Fig.~\ref{fig:xisham}).  

\begin{figure}[h]
\begin{center}
\includegraphics[bb=0 180 570 700, scale=0.35, angle=0]{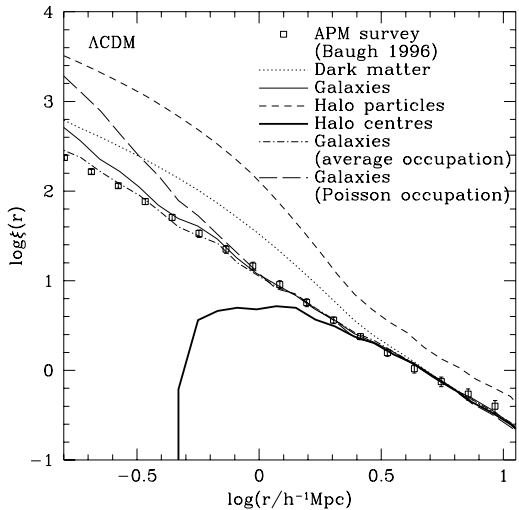}
\caption{Reproducing the clustering of galaxies in $\Lambda$CDM.
The correlation function of halos which contain galaxies is shown 
by the heavy solid line. This curve turns over below $r \sim 0.5 
h^{-1}$Mpc due to an exclusion effect which prevents halos 
overlapping. The correlation function of the dark matter 
particles in these haloes is shown by the long-dashed line; 
this puts too many pairs in massive haloes and leads to 
an overprediction of the small scale clustering. The number 
of galaxies predicted by a galaxy formation model set up to 
reproduce the luminosity function gives a reduced number of 
pairs by comparison with the particle case, and is in excellent 
agreement with the observed galaxy clustering. Based on a figure 
in \cite{Benson:2000}.}
\label{fig:xigal}
\end{center}
\end{figure}

So how can the power-law galaxy correlation function be understood, 
given the shape of the dark matter correlation function?  
\cite{Benson:2000} described the predictions of their galaxy 
formation model in these terms, and argued that a power-law could 
be obtained for the galaxy correlation function 
if the ``right'' number of galaxy pairs were predicted in each 
halo. Models which were set up to reproduce the galaxy 
luminosity function were found to predict a power-law 
galaxy correlation function in a $\Lambda$CDM cosmology. 
Fig.~\ref{fig:xigal} shows the components of the galaxy 
correlation function. The clustering of the halos occupied by 
galaxies is shown by the heavy solid line. Each halo has unit 
weight in this example. The curve turns over at small 
pair separations due to an exclusion effect; if halos got 
any closer to one another, they would be identified as a 
more massive halo by a percolation group finder. Considering 
only the dark matter particles contained within occupied 
dark matter haloes (long-dashed line) overpredicts the 
small scale clustering. The number of galaxy pairs within 
a halo does not increase with halo mass in proportion to 
the number of particles, so a lower clustering amplitude 
is predicted on small scales (light solid line). 

Today, these approaches have crystalized into two schemes: 
halo occupation distribution (HOD) modelling and sub-halo 
abundance matching (SHAM). 

\begin{figure*}
\begin{center}
\includegraphics[scale=0.65, angle=270]{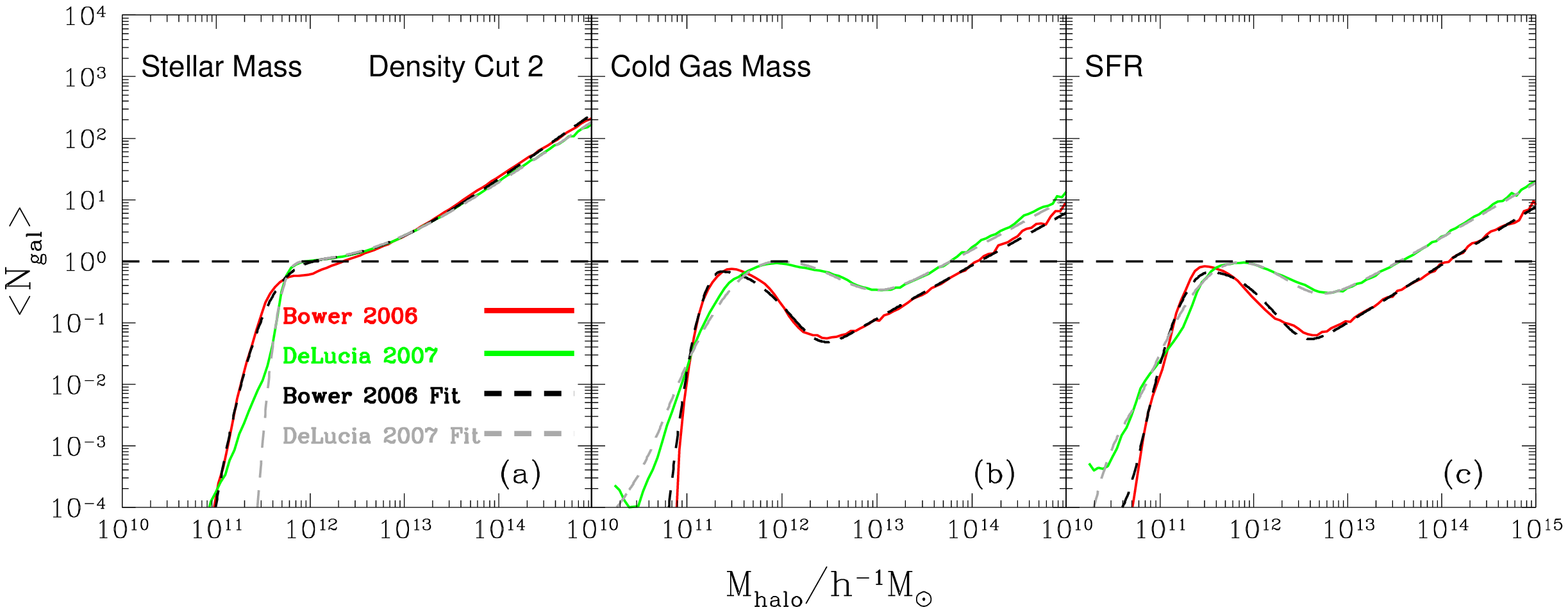}
\caption{The form of the halo occupation distribution {\it predicted} 
by two different semi-analytic galaxy formation models, the models of 
\cite{Bower:2006} and \cite{DeLucia:2007}. The HOD for galaxies 
selected according to a different intrinsic 
property is shown in each panel: left - stellar mass, middle - cold gas 
mass, right - star formation rate. In all cases, the samples have been 
ranked in terms of the intrinsic property, and the same abundance of 
objects is considered. The form of the HOD predicted for the cases of 
cold gas and star formation rate selected samples has a different form 
from that for stellar mass selected samples, with a peaked HOD for central 
galaxies. The dashed curves show how well parametric equations for the 
HOD can reproduce the forms predicted in the models. For stellar mass 
samples, a five-parameter fit gives a good match to the model results. 
For cold gas or star formation rate samples, a nine-parameter HOD is needed. 
Reproduced from \cite{Contreras:2013}.}\label{fig:HOD}
\end{center}
\end{figure*}

HOD modelling has its roots in the clump model of \cite{Neyman:1952}. 
In its modern form, HOD modelling took off around the start 
of the millennium, spurred on by the physical modelling described in the 
second part of this article. The HOD is a parametrization of 
the mean number of galaxies per halo. The HOD is split into contributions from central galaxies and satellite galaxies \cite{Zheng:2005}. Central galaxies are 
typically modelled using a softened step function, which encapsulates 
the transition from halos which are not massive enough to 
host a galaxy which meets the observational selection, to the mass 
for which all central galaxies are included. The mean number of 
satellite galaxies per halo is described by a power-law, which 
reaches unity at a higher halo mass than the central HOD \citep{Cooray:2002}.  
The canonical form used to model the HOD of optically selected galaxy 
samples is shown by the fit in the left panel of Fig.~\ref{fig:HOD}. 

A limitation of the HOD approach is that it is descriptive 
rather than predictive. Given an observational measurement 
of the clustering of galaxies, the parameters of the HOD can 
be constrained to reproduce this clustering, returning an 
interpretation of the measurement in terms of the number of 
galaxies per halo. The basic HOD machinery cannot make a prediction 
for a new clustering measurement, with for example, a different 
galaxy selection or at a different redshift. However, refinements 
to the HOD model to include galaxy luminosity and colour have been 
devised \citep{Skibba:2009}. 
As we will see later, the canonical form of the HOD outline above 
does not apply to all galaxy selections and there is no way 
to anticipate this without trying to implement a physical 
model of the galaxy population. Lastly, the basic assumption 
behind HOD modelling, that the clustering of dark matter 
haloes depends solely on halo mass has recently been 
demonstrated to be inaccurate \citep{Gao:2005,Croton:2006,Gao:2007,
Angulo:2009}. 

Sub-halo abundance matching (SHAM) is an even simpler 
approach to realising a galaxy distribution in an 
N-body simulation. The fundamental assumption behind 
SHAM is that there is a monotonic relation between 
a galaxy property, e.g. stellar mass, and the mass of the 
subhalo which hosts the galaxy. This relation is also 
assumed to have zero scatter. The subhalos from the simulation 
are then ranked in mass, breaking each halo into its 
component subhalos. A volume limited sample of galaxies, 
e.g. generated from a measurement of the galaxy 
luminosity function is then also ranked by the galaxy property 
(in this example, luminosity) 
and the two lists are paired off, with the most luminous  
galaxy being matched up with the most massive subhalo 
until the end of the list is reached \citep{Vale:2004,Conroy:2006}. 
In the simulation, the mass estimated for subhalos can 
be affected by stripping so the mass of the subhalo at infall 
is used in the SHAM procedure. 

SHAM seems to provide surprisingly good descriptions of 
observational samples \citep{Conroy:2006,Moster:2010}. This is 
all the more remarkable when one considers that no 
distinction is made regarding where the subhalo came from, 
that is, regardless of whether it was part of a cluster-mass 
halo or an isolated halo, there is assumed to be a connection 
to a galaxy property \citep{Watson:2013}. One might imagine that environmental factors 
would change the nature of the connection between subhalo mass 
and galaxy property for a satellite galaxy in a cluster. In 
SHAM, the subhalo mass is frozen at infall into a larger structure. 
Subsequently, the satellite galaxy could continue to form 
stars using up any available reservoir of cold gas, which 
would appear to change the subhalo mass - galaxy property 
relation. 

The SHAM approach has been extended to cope with the 
scatter in the galaxy property halo mass relation 
\citep{Moster:2010,Rodrigues-Puebla:2012}. The assumption which 
underpins SHAM has been evaluated by \cite{Simha:2012} 
using the output of a gas dynamic simulation. 
These authors found that the simulation produced relations 
between selected galaxy properties and subhalo mass which were 
monotonic, but with scatter. The scatter led to the clustering 
in a catalogue constructed by applying the SHAM hypothesis to 
differ somewhat from that in the original simulation output. 

The connection between empirical models of galaxy 
clustering based on the smoothed distribution of matter and 
those which start from haloes has recently been made 
\citep{Cacciato:2012}. In the next section we discuss 
a more physical approach which does not rely upon existing 
clustering data being available.

\section{Physical modelling of galaxy formation}

\begin{figure*}
\begin{center}
\includegraphics[bb=80 470 500 725, scale=1.0, angle=0]{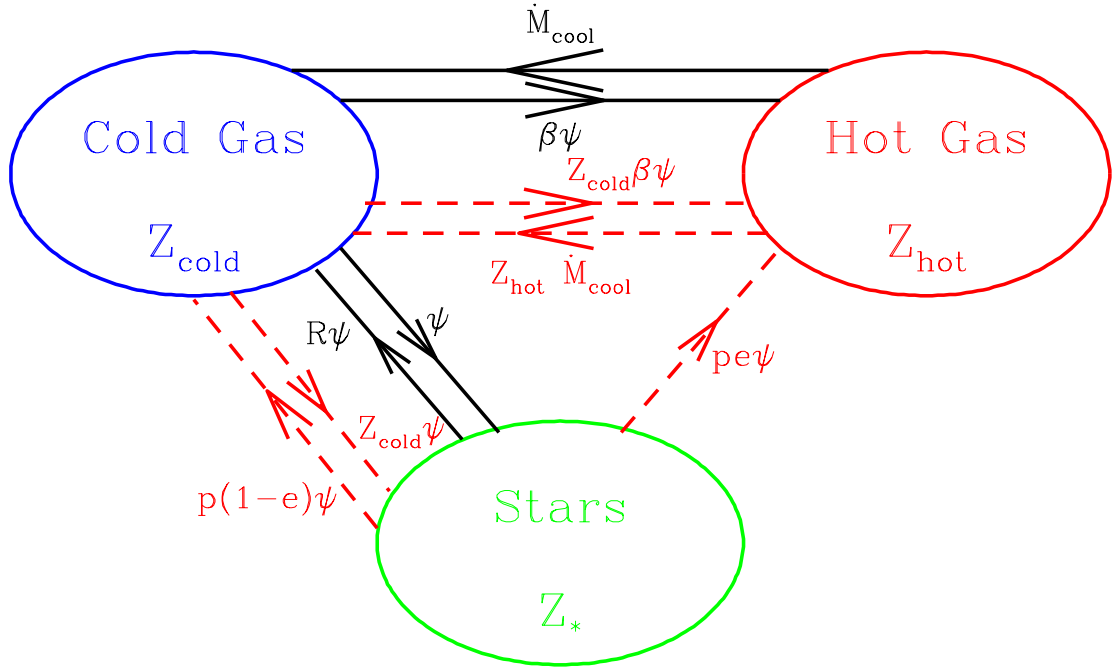}
\caption{The flow of mass and metals between reservoirs of hot gas, cold 
gas and stars. Semi-analytical models of galaxy formation solve the 
differential equations which describe the transfer of materials between 
these reservoirs. Reproduced from \cite{Cole:2000}.}
\label{fig:ISM}
\end{center}
\end{figure*}

By itself, the cold dark matter model says nothing directly about galaxy 
formation. Inferences can be drawn about the sequence of galaxy 
formation, based on how structures grow in the dark matter. 
However, without an attempt at a physically motivated calculation 
of the fate of baryons in a cold dark matter universe, there 
is little hope of learning much about galaxy formation or of 
understanding the implications of observations of high redshift 
galaxies for the cold dark matter cosmology (for reviews see 
Baugh 2006 and Benson 2010).  

\cite{White:1978} argued that galaxy formation is a two-phase 
process, with the bulk of the mass undergoing a dissipationless 
collapse which is responsible for building the gravitational 
potential wells or halos in which galaxies form. The baryonic 
component of the universe is able to dissipate 
energy, and therefore to collapse down to smaller scales, forming 
denser units, which retain their identity within the cluster. 
This model was able to explain the appearance of 
clusters of galaxies. However, without an additional process to 
reduce the efficiency of galaxy formation in shallow gravitational 
potential wells, the predicted luminosity function is much steeper 
than is observed at the faint end.  

This pioneering work, along with a clutch of papers published around 
the same time looking at the radiative cooling of gas within gravitational 
potential wells, laid the groundwork for modern galaxy formation 
theories. The break in the galaxy luminosity function can be understood 
by comparing the time taken for gas to cool to the age of the universe. 
The time taken for all of the gas within a halo 
to cool radiatively increases with halo mass. This is because 
cooling is a two-body process (collisionally excited radiative 
transitions or bremsstrahlung) which depends 
on the square of the gas density. In hierarchical models, more 
massive haloes tend to form later when the density of the universe is lower. 
It is possible for the cooling time of the gas to exceed the Hubble time, 
thus limiting the supply of cold gas to form a galaxy 
(see the review of Fred Hoyle's contributions to galaxy 
formation theory by Efstathiou 2003). 

The first papers to incorporate these ideas fully into the cold dark 
matter cosmology, introducing the semi analytical 
methodology, were published in 1991 \citep{White:1991,Cole:1991,Lacey:1991}. 
This approach tries to follow a wide range  
of the processes which are thought to be important in determining 
the fate of the baryons. This is a daunting task. At the time, theories 
of star formation were rudimentary at best. There has been much 
progress in this area since 1991, but we are still a long way 
from having a reliable description of the process which underpins 
galaxy formation. The regulation of star formation efficiency comes 
from the stars themselves. 
Stars above $\approx 5-8$ times the mass of the 
Sun end their life in a Type II supernova, which injects substantial amounts 
of energy and momentum into the interstellar medium. 
This alters the state of the gas in the interstellar medium (ISM), 
perhaps leading to the ejection of gas 
from the galactic disk or even the dark matter halo. 
This process is known as supernova feedback and is critical to the 
success of any model of galaxy formation. 

The absence of a precise description of a key process, such as 
star formation and supernova feedback, may lead one to consider 
giving up any hope of ever understanding galaxy formation. Instead, 
in semi analytical modelling an attempt is made to write down 
the differential equation which gives the current best bet model 
of how the system behaves. As our understanding develops, or when 
new observations clarify how a process works, then the model can be 
improved. The differential equation may contain a free parameter. 
Often there is little guidance as to the appropriate range of values 
to take for the parameter. 
In such instances, the only approach is to be pragmatic and 
see what the model predicts for different parameter values. By comparing 
the model predictions to observations, the value of the parameter is set 
as the one which gives the most faithful reproduction of the data. 
This procedure is exactly what physicists undergo when attempting to 
describe complex phenomena: start off with a simple model, which can 
be adjusted or refined to improve the description of the observations. 
I will give an example of this principle in action in the next section. 

The semi analytical framework allows us to model a range of processes 
together, within the cosmological setting of the formation of 
structure in the dark matter. The ability to follow the interplay 
between processes is essential in studying galaxy formation. The models 
solve the set of differential equations which govern the flow of 
mass and metals between different reservoirs of baryons: hot gas, 
cold gas and stars (Fig.~\ref{fig:ISM}). The output of the models is the full star formation 
and chemical enrichment histories for a wide range of galaxies, including 
mergers between galaxies. 

Semi analytical modelling has some features which might be perceived 
as limitations or drawbacks. One example is the generality of the  
assumptions which are needed to be able to calculate the fate of 
the baryonic component. 
Another is the ``deterministic" way in which 
processes such as supernova feedback are modelled. In the 
semi analytical model, the mass loading of the supernova 
driven wind is specified by choosing model parameters, and 
precisely this amount of gas is ejected from the ISM. 
In a gas dynamics simulation in which the wind is fully 
coupled to the hydrodynamics equations (note this is not generally 
the case, with a semi-analytical model of feedback inserted into 
the simulation to describe feedback), the same number of 
supernovae could result in a very different mass of gas 
being ejected. The mass loading could be intricately linked 
to the resolution of the simulation.  

Nevertheless, despite the progress made over the past twenty years, 
there is still widespread mistrust of semi analytical modelling. 
This has led to a burgeoning reductionist movement in galaxy 
formation in which simplified models have been devised with the aim 
of elucidating how galaxies form. Examples include the ``bathtub" 
and ``reservoir" models \citep{Bouche:2010,Dave:2012}. 
These calculations are inspired by 
models of supply and demand from economics, and track the 
inflow (sources) of gas into halos and the ``sinks" of cold gas in 
star formation and supernova feedback. In their simplest form, 
the models follow one galaxy per halo, and invoke ad-hoc efficiency 
factors to specify the inflow of gas as a function of halo mass, 
without any attempt to calculate the rate at which gas can cool 
or to explain the form of the efficiency factor. 
Galaxy mergers are ignored. This class of calculation effectively 
takes one of the equations which has been considered within 
semi analytical models for more than two decades and solves it in 
isolation. 

The desire for a better grasp of how galaxy properties 
are shaped by different processes is understandable, but it is 
not clear that it can be usefully gained from such stripped-down 
approaches. The perceived ``complexity" of semi analytical 
modelling is actually the great strength of the technique. 
The ability to model the interplay between processes is the key 
to building a realistic model of galaxy formation. 
By taking a more complete view of galaxy formation rather 
than a selective one, the consequences of the calculation - the 
predictions of the model - are more far reaching and therefore 
more tightly constrained by observations. If the model seems complex, 
then this is simply a reflection of the nature of the underlying 
processes, such as star formation and heating by supernovae.

Semi-analytical modelling of galaxy formation is complementary to 
the approach of using a gas dynamics simulation, with the two techniques 
having many aspects in common. In general, gas dynamics simulations 
rely on fewer assumptions to follow some of the processes in 
galaxy formation. For example, the treatment of gas cooling in 
semi-analytical models assumes spherical symmetry, whereas this is 
not necessary in a hydrodynamics simulation. Nevertheless, in carefully 
controlled comparisons, the modelling of gas cooling in semi-analytical 
models can produce the same results that are obtained in the 
hydro-simulation \citep{Yoshida:2002,Helly:2003,DeLucia:2012}.
In other areas, the two methods are more similar than many people 
realize. A good illustration is star formation, which is firmly ``sub grid" 
in simulations which aim to follow more than one galaxy. The treatments 
of star formation in a gas dynamics simulation and in a semi-analytical 
model are very similar. Further discussion of how star formation is treated 
in semi-analytical models is given in the next section. A key limitation 
on the use of gas dynamics simulations to model galaxy clustering is their 
computational expense and the requirement for ``sufficient" resolution in 
mass and length scales \citep{Governato:2007}. These considerations have 
tended to force gas simulators to use relatively small simulation boxes, 
typically measured in tens of megaparsecs. This in turns limits the 
predictions for the clustering to pair separations of a few magaparsecs. 
An alternative to trying to predict the galaxy correlation function is 
to focus instead on hhow haloes are populated with galaxies. If enough 
different environments can be sampled, e.g. by resimulating patches from a 
larger volume at high resolution and with gas \citep{Crain:2009}, 
then such a simulation could be used to predict the halo occupation 
distribution. One advantage of gas simulations over semi-analytics is that 
they can follow the redistribution of matter due to outflows of baryons. 
Calculations using the Over Whelmingly Large Simulations have shown 
that the physics of galaxy formation, particularly AGN feedback, 
has an impact on the distribution of matter which has implications 
for the interpretation of weak lensing measurements 
\citep{Semboloni:2011,vanDaalen:2011}. 

Nevertheless to address clustering on scales of tens to hundreds of 
megaparsecs, the only viable technique is semi-analytics used in 
combination with large volume, high resolution N-body simulations 
of the clustering of dark matter, which we focus on in the later 
sections of this review.

\begin{figure*}
\begin{center}
\includegraphics[scale=0.65, angle=0]{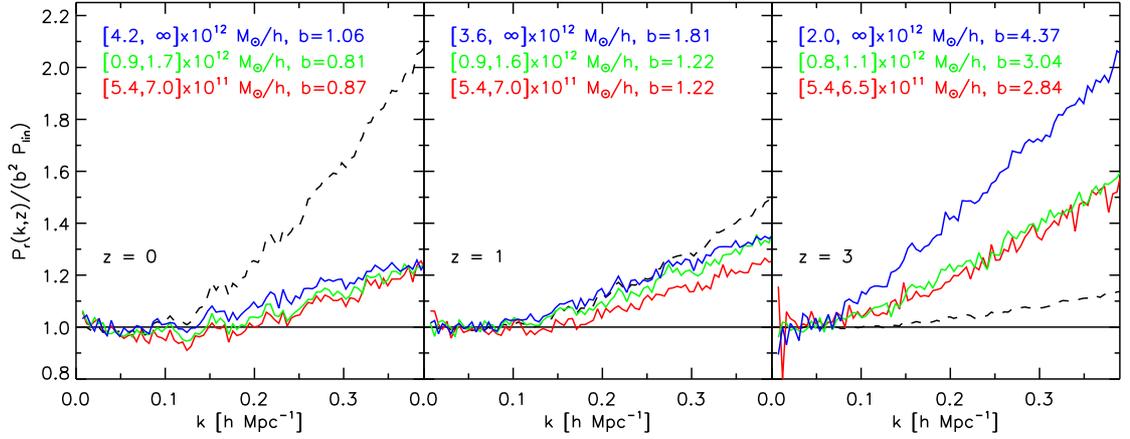}
\caption{The scale dependent bias of haloes of different 
mass, as measured from a very large volume N-body 
simulation. Each panel corresponds to a different 
redshift as labelled. The halo mass range and the 
measured asymptotic bias are given by the legend. If 
the asymptotic bias described the halo power spectrum, 
the ratio of the halo power spectrum divided by the 
linear power spectrum multiplied by the square of this 
bias would be unity. The clustering of haloes measured 
from the simulation deviates strongly from a ratio of unity, 
which indicates that the halo bias is scale dependent. 
Furthermore, the shape of these curves is different 
from that corresponding to the nonlinear matter power 
spectrum divided by the linear theory spectrum (shown 
by the dashed line). Reproduced from \cite{Angulo:2008}.
}\label{fig:bhalo}
\end{center}
\end{figure*}

\begin{figure*}
\begin{center}
\includegraphics[scale=0.65, angle=0]{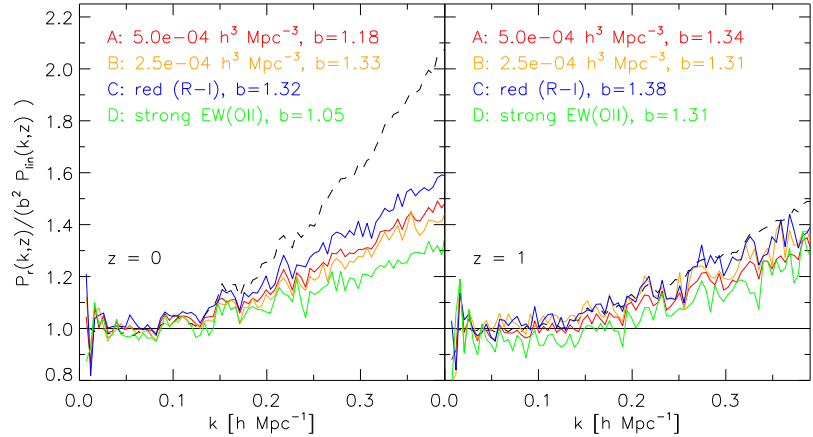}
\caption{
The predicted scale-dependent bias in the galaxy 
distribution. As in the previous figure, the power 
spectrum measured for different galaxy 
selections is divided by the linear theory 
power spectrum multiplied by the square of the 
asymptotic bias. Different colours correspond to 
different selections: red and orange show the predicted 
clustering for flux limited samples, the blue curves show 
the power spectrum for red galaxies and the green curves 
show galaxies with strong emission lines. Reproduced from 
\cite{Angulo:2008}.}
\label{fig:bgal}
\end{center}
\end{figure*}

\section{Illustration: the 
star formation rate in galaxies} 

An illustration of how semi analytical models work can be 
obtained by considering recent progress in how star formation 
is modelled within a galaxy. 

The bulk of semi analytical models attempt to predict the global 
star formation rate within a galaxy. The early modelling of the 
star formation rate was essentially based on dynamical arguments, 
with loose motivation coming from a comparison to the Kennicutt-Schmidt 
law \citep{Bell:2003}. 
The star formation rate, $\psi$, is often parametrized as 

$$
\psi = \epsilon \frac{M_{\rm cold}}{\tau}, 
$$ 
where  $M_{\rm cold}$ is the total mass of cold gas in the galaxy 
and $\epsilon$ is an efficiency factor which controls 
the fraction of cold gas which is turned into stars in the  
timescale $\tau$. The timescale for star formation is 
generally assumed to scale with the dynamical time within 
the galaxy: 

$$
\tau = t_{\rm dyn} f (v_{\rm disk}). 
$$ 
In some models, $f(v_{\rm disk}) = 1$; in the \cite{Cole:2000} 
model, an explicit scaling of the star formation timescale with the 
circular velocity of the disk was implemented, to allow the model 
to produce a better match to the observed gas fraction luminosity 
relation for spiral galaxies: $f(v_{\rm disk}) = \left(v_{\rm disk}/200 
{\rm km s}^{-1}\right)^{\alpha_*}$. Hence in the most general case, 
two parameters are required to set the star formation rate: $\epsilon$ 
and $\alpha_*$. These parameters are set by chosing values, running 
the model and then comparing the model predictions to observables. The 
key observables for constraining the values of these star formation 
parameters are the gas fraction - luminosity relation, the galaxy 
luminosity function and the colour magnitude relation. 

High resolution imaging of galaxies at different wavelengths has revealed 
that star formation activity correlates better with the molecular hydrogen 
content of galaxies than with the overall cold gas mass. \cite{Lagos:2011a} 
investigated more general star formation models  in the {\tt GALFORM} 
semi analytical model, implementing different 
empirical and theoretically motivated star formation laws 
(see also Cook et~al. 2010 and Fu et~al. 2010). 
The most successful of these was the empirical star formation law 
proposed by \cite{Blitz:2006}, who suggested that the 
observational data could be explained if the ratio of molecular to 
atomic hydrogen is set by the pressure in the mid-plane of 
galactic disks; gas disks with higher pressure have a higher 
fraction of H$_{2}$.  

This work illustrates the modularity of semi analytical modelling and 
how it provides a framework in which new and improved descriptions of 
various processes can be readily implemented. The Blitz \& Rosolowsky 
star formation law 
involves two observationally determined ``parameters". Whereas in the 
original parameterization of the star formation rate there was little guidance 
about the range of parameter values which should be considered, there 
is now a much smaller volume of parameter space to search (at least once 
the Blitz \& Rosolowsky law has been adopted). Furthermore, as the 
modelling of the star formation becomes more sophisticated, the predictions 
that can be made by the model expand. Rather than simply outputting 
the cold gas mass of galaxies, the atomic and molecular hydrogen 
contents are now predicted, meaning that the model should also be able 
to reproduce the mass functions of HI and H$_{2}$, their evolution and 
their relation to other galaxy properties \citep{Lagos:2011b}. 
By combining {\tt GALFORM} with the photon dominated region model 
of \cite{Bell:2006}, it is also possible to predict the 
different carbon monoxide transitions, and to make contact with 
observations from ALMA \citep{Lagos:2012}. 

Hence by adopting the improved star formation model, the parameter space 
open to the model has shrunk in volume and the constraints on the model 
have increased through the capability to make new predictions which 
must match the available observations. 

\section{Predictions for galaxy clustering}

The combination of a semi-analytical model of galaxy 
formation with a cosmological N-body simulation extends 
the capability of the models to make predictions for 
the spatial distribution of galaxies \citep{Kauffmann:1999,
Benson:2000}. The models follow the physics of the baryonic 
component of the universe to predict how many galaxies populate 
dark matter haloes as a function of their mass and formation 
history, and tells us the properties of these galaxies. The 
semi-analytical model therefore predicts the mean number of 
galaxies per halo and it was the description of the model 
output in these terms which helped to stimulate the 
development of HOD modelling. 

The form of galaxy bias can be understood by first looking 
at the clustering of dark matter haloes. The canonical model 
is that the clustering of halos can be described by multiplying 
the matter power spectrum by the square of an asymptotic bias factor. 
Formally, the bias factor should be applied to the linear power 
spectrum of matter fluctuations. 
\cite{Angulo:2008} investigated this hypothesis with 
a moderate resolution N-body simulation of a very large cosmological 
volume, measuring $1340 h^{-1}$Mpc on a side. Fig.~\ref{fig:bhalo} 
shows the ratio of the power spectrum measured for different samples 
of dark matter haloes divided by a scaled linear theory power spectrum. 
The scaling is the square of the asymptotic bias, which is measured from  
the simulation on very large scales (small $k$). This ratio deviates 
strongly from unity at quite large scales, typical of those used to 
fit BAO. This means that a simple bias squared times the linear theory 
spectrum is not a good way to describe halo clustering. 
If the linear power 
spectrum is replaced by the nonlinear matter power spectrum in 
the simulation, there is some improvement, but there are still substantial 
deviations, as shown by the discrepancy between the coloured curves 
and the dashed black line in Fig.~\ref{fig:bhalo}. This disagreement is 
particularly strong at high redshift, where the resolved haloes correspond 
to higher peaks in the density field than they do at lower redshifts. 

The next step in the calculation is to combine the large volume 
N-body simulation with a semi-analytical model of galaxy formation. 
This is the only way to make predictions for galaxy clustering on 
scales of tens of megaparsecs and above. Current simulations which 
follow the hydrodynamics of the gas are restricted to volumes which 
are several thousand times smaller, and can only reliably predicted 
galaxy clustering out to pair separations of a few megaparsecs. 
Fig.~\ref{fig:bgal} reveals that both the asymptotic bias and 
the form of the scale dependence of the bias depend upon how 
galaxies are selected \citep{Angulo:2008}. This in turns has implications 
for the apparent positions of the BAO when observed using different 
galaxy tracers. 

Finally one might ask, given the uncertainty in the modelling of 
the processes behind galaxy formation, how far can we trust the 
predictions of semi-analytical models for galaxy clustering? The 
Millennium N-body simulation of \cite{Springel:2005} provides an 
excellent test-bed on which different semi-analytical models can 
be run and compared. \cite{Contreras:2013} compared 
the clustering predictions of the Durham and Munich models 
\citep{Bower:2006,DeLucia:2006,Bertone:2007,Font:2008,Guo:2011}. 
These groups have developed independent models which follow the same 
processes but with different implementations. These differences 
even extend to the first step in the galaxy formation code of 
extract merger histories for dark matter halos from the simulation. 
A summary of the comparison is given in Fig.~\ref{fig:HOD}. 
The different models give remarkably similar predictions for 
the HOD (an {\it output} of the models) for galaxy samples 
selected by stellar mass. The results are qualitatively similar 
for samples selected by the cold gas mass or star formation 
rate of the galaxies, but differ in detail. These differences 
can be traced to the way in which star formation is modelled 
by the different groups. 

\section{Conclusions}

I have discussed empirical and physical methods for connecting 
dark matter haloes to galaxies. Empirical methods include: 1) 
Applying a weighting scheme to the smoothed dark matter density 
field. 2) Applying a weighting of dark haloes through the HOD which 
specifies the mean number of galaxy pairs as a function of halo 
mass. 3) SHAM, in which galaxies and subhalos are first ranked 
and then matched up. The physical approach is to carry out a 
calculation of the fate of baryons in a cold dark matter universe 
to predict which galaxies are in which haloes. Currently, this 
is only possible in cosmologically representative volumes by using 
a semi-analytical model of galaxy formation. I briefly reviewed 
how these models work and gave an illustration of the power of this 
approach by discussing recent work on improved models of the  
star formation rate in galaxies.  

Much progress has been made in understanding the connection between 
haloes and galaxies and hence of galaxy bias. One clear conclusion  
so far is that galaxy bias is {\it scale dependent} and depends 
sensitively on the selection applied to construct the sample. This 
needs to be taken into account when analysing large-scale structure 
as a cosmological probe so that all of the data can be utilized. 
A comparison of the predictions from different models which aim to follow 
the same processes in galaxy formation gives some encouraging results 
\citep{Contreras:2013}. 
The predictions for samples selected by stellar mass seem robust. However, 
there is more discrepancy between the predictions for other galaxy 
selections which are closer to what will be used in future galaxy 
surveys. This suggests that further theoretical work is needed if we 
are to maximize the potential of future surveys to tells us the values 
of the basic cosmological parameters and about the physics of galaxy formation.

\section*{Acknowledgments} 
I would like to thank the organizers for inviting me to 
speak at such a stimulating meeting and for being indulgent 
in extending the deadline for this contribution. 

%\begin{thebibliography}{}
%\end{thebibliography}

\bibliographystyle{mn2e}
\bibliography{Biblio}

%\end{multicols}

\end{document}